\documentclass [12pt] {article}
\usepackage{color,graphicx,amssymb,amsmath}
\usepackage{pstricks,pst-node}
\usepackage[all]{xy}

\newcommand{\makebf}[1]{{\mbox{\bf\boldmath ${#1}$}}}
\newcommand{\makebfscript}[1]{{\mbox{\bf\boldmath\scriptsize ${#1}$}}}
\newcommand{\bfa}{{\makebf{a}}}
\newcommand{\bfb}{{\makebf{b}}}
\newcommand{\bfscripta}{{\makebfscript{a}}}
\newcommand{\bfscriptb}{{\makebfscript{b}}}
\newcommand{\bff}{{\makebf{f}}}
\newcommand{\bfg}{{\makebf{g}}}
\newcommand{\bfp}{{\makebf{p}}}
\newcommand{\bfr}{{\makebf{r}}}
\newcommand{\bfx}{{\makebf{x}}}

\newcommand{\bfy}{{\makebf{y}}}
\newcommand{\bfi}{{\makebf{i}}}
\newcommand{\bfj}{{\makebf{j}}}

\newcommand{\calC}{{\cal C}}
\newcommand{\calL}{{\cal L}}

\newcommand{\calP}{{\cal P}}

\newcommand{\calQ}{{\cal Q}}
\newcommand{\bfgy}{\bfg_{\mbox{\scriptsize $\bfy$}}}
\newcommand{\gy}{g_{\mbox{\scriptsize $\bfy$}}}
\newcommand{\bfgi}{\bfg_{\mbox{\scriptsize $\bfi$}}}
\newcommand{\bfgj}{\bfg_{\mbox{\scriptsize $\bfj$}}}

\newtheorem{theorem}{Theorem}
\newtheorem{lemma}{Lemma}
\newtheorem{example}{Example}
\newtheorem{definition}{Definition}

\begin{document}

\title{A ``No-Go'' Theorem for the Existence of an Action Principle for Discrete Invertible Dynamical Systems}
\author{
Gianluca Caterina and Bruce Boghosian\\
{\scriptsize\it Department of Mathematics, Tufts University, Medford, Massachusetts 02420, USA}
}
\date{\today}
\maketitle

\begin{abstract}
In this paper we study the problem of the existence of a least-action principle for invertible, second-order dynamical systems, discrete in time and space. We show that, when the configuration space is finite, a least-action principle does not exist for such systems.  We dichotomize discrete dynamical systems with infinite configuration spaces into those of {\it finite type} for which this theorem continues to hold, and those not of finite type for which it is possible to construct a least-action principle.  We also show how to recover an action by restriction of the phase space of certain second-order discrete dynamical systems.  We provide numerous examples to illustrate each of these results.
\end{abstract}

\section{Introduction}

Hamilton's Principle of Least Action provides a unifying and elegant reformulation of classical mechanics.  If we suppose that the instantaneous state of a mechanical system is described by generalized coordinates $\bfr$ taking values in a configuration space $\calC$, then Hamilton's Principle states that the trajectory of the system between times $0$ and $T$ will be such as to extremize the {\it action} functional
\begin{equation}
S[\bfr] = \int_{0}^{T} dt\; \calL(\dot{\bfr},\bfr),
\label{eq:action}
\end{equation}
where the {\it Lagrangian} $\calL:T\calC\rightarrow{\mathbb R}$ is a real-valued function on the tangent bundle of $\calC$, and where the dot denotes differentiation with respect to time $t$.  Setting the Fr\'{e}chet derivative of $S[\bfr] $ to zero,
\begin{equation}
0=\frac{\delta S[\bfr] }{\delta \bfr(t)} = \frac{\partial\calL}{\partial\bfr} - \frac{d}{dt}\left(\frac{\partial\calL}{\partial\dot{\bfr}}\right),
\end{equation}
yields Lagrange's equations of motion for the mechanical system.

\begin{example}
A mechanical system with mass matrix $M$ and potential energy function $V:\calC\rightarrow{\mathbb R}$ has the Lagrangian function
\begin{equation}
\calL(\dot{\bfr},\bfr) = \frac{1}{2}\dot{\bfr}^T M\dot{\bfr} - V(\bfr),
\end{equation}
where the $T$ superscript denotes transpose.  Lagrange's equations of motion for this system are easily seen to be equivalent to Newton's equations
\[
M\ddot{\bfr} = -\frac{\partial V}{\partial\bfr}.
\]
\end{example}

Newton's equations are second-order autonomous ordinary differential equations that are well known to be {\it reversible}; that is, they are invariant under time reversal, $t\rightarrow -t$.  They can be reformulated as a system of first-order differential equations known as Hamilton's canonical equations,
\begin{eqnarray}
\dot{\bfr} =  \frac{\partial H}{\partial\bfp}=M^{-1}\bfp \\
\dot{\bfp} =  -\frac{\partial H}{\partial\bfr}=  -\frac{\partial V}{\partial\bfr},
\end{eqnarray}
where we have introduced the Hamiltonian $H=\frac{1}{2}\bfp^{T}M^{-1}\bfp+V(\vec{\bfr})$ and the momentum $\bfp$.  Under time reversal the momentum changes sign, $\bfp\rightarrow -\bfp$, and the time derivatives change sign, $d/dt\rightarrow -d/dt$, so Hamilton's canonical equations are invariant.

All of the above is well known for mechanical systems with continuous configuration spaces $\calC$, and with a continuous time coordinate $t$.  If time is discrete, as it is in a finite-difference algorithm, the configuration of the system at time $t\in{\mathbb Z}$ may be denoted by $\bfr(t)$, and the trajectory of the system from time $0$ to time $T\in{\mathbb Z}^+$ is given by the sequence $\bfr\equiv\{\bfr(t)\, |\, 0\leq t < T\}$.  In this case, we might suppose that the action functional in Eq.~(\ref{eq:action}) could be recast as a function of $\bfr$,
\begin{equation}
S(\bfr) = \sum_{t=0}^{T-1} \calL\left(\bfr(t),\bfr(t+1)\right),
\label{eq:discreteaction}
\end{equation}
where $\calL:\calC^2\rightarrow{\mathbb R}$ is the discrete-time analog of the Lagrangian function.  The equations of motion may be obtained by setting to zero the gradient of the above,
\begin{equation}
0 = \frac{\partial S(\bfr)}{\partial\bfr(t)} =\
\calL_2(\bfr(t-1),\bfr(t)) +
\calL_1(\bfr(t),\bfr(t+1)),
\label{eq:discreteLagrangian}
\end{equation}
where $\calL_1$ and $\calL_2$ denote the gradient of $\calL$ with respect to its first and second arguments, respectively.  If $\calL$ is a symmetric function of its two arguments, then $\forall_{{\bfscripta,\bfscriptb}\in\calC}: \calL_1(\bfa,\bfb)=\calL_2(\bfb,\bfa)$, from which it quickly follows that Eq.~(\ref{eq:discreteLagrangian}) is invariant if $t-1$ and $t+1$ are interchanged, demonstrating reversibility of the discrete-time mechanical system.  This formulation of discrete-time mechanics seems to have been first posited by Baez and Gilliam~\cite{bib:baez}.  More recently, it has recently been demonstrated~\cite{bib:marsden} that this approach may be used to construct symplectic finite-difference algorithms for the numerical simulation of mechanical systems.

The utility of action principles for mechanical systems in discrete time with a finite discrete set of states, such as cellular automata, has received much less attention.  We might still suppose that Eq.~(\ref{eq:discreteaction}) is a reasonable definition for the action, but it must be understood that $\bfr(t)$ takes its values in a discrete configuration space $\calC$ of finite cardinality, such as a finite set of integers.  One immediately runs into the difficulty that it is impossible to take derivatives of $\calL$ with respect to its two arguments, so $\calL_1$ and $\calL_2$ are not well defined.  Baez and Gilliam addressed this problem using methods of algebraic geometry~\cite{bib:baez}, but here we take a different approach.  Rather than attempt to salvage Eq.~(\ref{eq:discreteLagrangian}), we instead give primacy to Eq.~(\ref{eq:discreteaction}).  For a discrete configuration space $\calC$, we pose the question of whether or not it is possible to find a real-valued function $\calL:\calC^2\rightarrow{\mathbb R}$ such that the action $S(\bfr)$ given by Eq.~(\ref{eq:discreteaction}) is smallest for trajectories governed by a reversible evolution rule for the finite-state system-- one giving $\bfr(t+1)$ in terms of $\bfr(t)$ and $\bfr(t-1)$ that is invariant under interchange of $t+1$ and $t-1$.  In this paper, we answer this question in the negative.  Indeed, we disprove the existence of an action principle in a finite configuration space not only for the case of reversible dynamical systems, but also for the much more general case of invertible dynamical systems.  The proof is general enough to apply to any invertible dynamical system with a finite numbers of states, including all invertible cellular automata.

Our approach is based on an observation of Feynman~\cite{bib:feynman} who noted that, if an action is minimized on a physical path, then it must be minimized on every ``infinitesimal subsection'' of that path.  Because our system evolves in discrete time steps, we replace the notion of ``infinitesimal subsection'' with its discrete counterpart, examining the existence of action principles for a sequence of three time steps.  This approach leads to a set of inequalities that must be satisfied by the Lagrangian function, and we show that these inequalities are necessarily inconsistent for any invertible dynamical system with a finite number of states.

Because this paper is meant to be self-contained, we begin by reviewing the notions of reversibility and invertibility for discrete dynamical systems.  We then describe a representation of such systems via elements of the symmetric group; this representation will play an important role in our subsequent proof.  Finally, we make precise the concept of action for a finite-state dynamical system and we present a proof of the ``No-Go Theorem'' for an action principle. 

We conclude the paper by observing that, if we relax the hypothesis of finiteness of the configuration space or if we work on a suitable restriction of the phase space, an action principle can be recovered. Two examples describing how an action can be recovered in these cases are discussed in the conclusions section of our work.

\section{Discrete Dynamical Systems}

We consider dynamical systems that are discrete in time and discrete in state.  Discreteness in time means that the evolution of the system takes place in successive time steps.  Discreteness in state means that at each time step the system may be in only one of $r\in {\mathbb Z}^+$ states; we refer to the set of all possible instantaneous states of the system as the {\it configuration space}, $\calC\equiv\{0,1,\ldots,r-1\}$.  (Later in the paper, we consider the limit of infinite configuration space, $r\rightarrow\infty$.)  The state of the system at time $t$ will be denoted by $\bfr(t)$.

\begin{example}
The instantaneous state of many discrete dynamical systems, such as cellular automata, is most conveniently described by $n$ dependent variables, each of which takes its values in a set $\calQ$, with $|\calQ|=q$.  This is equivalent to describing the system by a single state variable which takes values in a configuration space $\calC=\calQ^n$ of size $r=q^n$.  In this case, we denote the $j$'th dependent variable at time step $t$ by $r_j(t)\in\calC$, for $0\leq j<n$ and $t\geq 0$, so that $\bfr(t)$ may be thought of as an $n$-vector.
\end{example}

We refer to a dynamical system as {\it $p$'th order} if the state at time $t$ depends on those at times $t-1,t-2,\ldots,t-p$.  In this paper, we shall focus on the case $p=2$.  A first-order ($p=1$) dynamical system has an evolution rule of the form
\begin{equation}
\bfr(t+1) = \bff\left(\bfr(t)\right),
\label{eq:fo}
\end{equation}
where we have introduced the function $\bff:\calC\rightarrow\calC$.  A second-order ($p=2$) dynamical system has an evolution rule of the form
\begin{equation}
\bfr(t+1) = \bfg\left(\bfr(t),\bfr(t-1)\right).
\label{eq:so}
\end{equation}
where we have introduced the function $\bfg:\calC^{2}\rightarrow\calC$.  More generally, a $p$'th-order dynamical system requires the introduction of a map $\calC^{p}\rightarrow\calC$.  A $p$'th-order dynamical system has a well-posed initial-value problem if $\bfr$ is specified at $p$ successive initial times, say from $t=0$ to $t=p-1$. In what follows, we denote by $\Gamma$ the dynamical system induced by $\bfg.$

\subsection{Configuration Space, Phase Space, and Periodic Orbits}

For deterministic first-order dynamics, $p=1$, it is evident that the system will return to an initial configuration in at most $|\calC|=r$ time steps, after which the motion repeats.  For $p>1$, however, the evolution in configuration space is not manifestly Markovian, since the configuration of the system at time $t+1$ depends on that at $p$ earlier time steps.

To recast the dynamics as manifestly Markovian, it is necessary to consider the set of all $p$-tuples of configurations.  Each such $p$-tuple shall be called a {\it phase point}, and the set of all of them will be called the {\it phase space}, $\calP=\calC^p$.  The advantage of this approach is that the $p$-tuple at time $t+1$ depends only on that at time $t$; that is, the phase-space dynamics are first-order.

The system will return to an initial phase point in at most $|\calP|=|\calC|^p=r^{p}$ time steps, after which point the motion repeats.  The number of time steps required to return to an initial phase point is called the {\it recurrence time} and denoted by $\tau$.  Generally, the recurrence time will depend on both the dynamical rule and the initial condition~\cite{bib:coppersmith}.

\subsection{Invertibility and Reversibility}

A $p$'th-order dynamical system is said to be {\it invertible} if it is possible to uniquely determine $\bfr(t-p)$ given $\bfr(t-p+1),\ldots,\bfr(t)$.  For a first-order dynamical system, invertibility means that it must be possible to solve Eq.~(\ref{eq:fo}) for $\bfr(t)$ or, in other words, that $\bff$ must be an invertible function.  For a second-order dynamical system, this means that it must be possible to solve Eq.~(\ref{eq:so}) for $\bfr(t-1)$.  Thus, an invertible dynamical system admits a rule for evolving backwards in time, as well as forwards in time.  Alternatively stated, it has a well posed final-value problem, if $\bfr$ is specified at $p$ successive final times, say from $t=T-p+1$ to $t=T$.

An invertible dynamical system is said to be {\it reversible} if the rule for evolving backwards in time is identical in form to that for evolving forwards in time~\cite{bib:toffoli}.  This is substantially more restrictive than invertibility.  For the first-order ($p=1$) dynamical system in Eq.~(\ref{eq:fo}), it means that 
\begin{equation}
\bfr(t) = \bff\left(\bfr(t+1)\right).
\label{eq:forev}
\end{equation}
For the second-order ($p=2$) dynamical system in Eq.~(\ref{eq:so}), it means that 
\begin{equation}
\bfr(t-1) = \bfg\left(\bfr(t),\bfr(t+1)\right).
\label{eq:sorev}
\end{equation}
These requirements place severe restrictions on the form of the functions $\bff$ and $\bfg$.  For example, in the first-order ($p=1$) case, we can combine Eqs.~(\ref{eq:fo}) and (\ref{eq:forev}) to find
\begin{equation}
\bfr(t) = \bff\left(\bff\left(\bfr(t)\right)\right).
\end{equation}
Now any possible $\bfr(t)\in\calC$ could be specified, perhaps as an initial condition, and the above must hold for any of them, so we require
\begin{equation}
\forall_{\bfx\in\calC}: \bfx = \bff\left(\bff\left(\bfx\right)\right).
\label{eq:frequire}
\end{equation}
Thus, any reversible, first-order, discrete-time dynamical system must have a recurrence time of at most two.  This makes them substantially less interesting than reversible dynamical systems of higher order.

In the second-order ($p=2$) case, we can combine Eqs.~(\ref{eq:so}) and (\ref{eq:sorev}) to find
\begin{equation}
\bfr(t-1) = \bfg\left(\bfr(t),\bfg\left(\bfr(t),\bfr(t-1)\right)\right).
\end{equation}
Again, any possible 2-tuple $\{\bfr(t-1),\bfr(t)\}\in\calP=\calC^2$ could be specified, perhaps as an initial condition, and the above must hold for any of them, so we require
\begin{equation}
\forall_{\bfx,\bfy\in\calC}: \bfx = \bfg\left(\bfy,\bfg\left(\bfy,\bfx\right)\right).
\label{eq:grequire}
\end{equation}

The requirements expressed by Eqs.~(\ref{eq:frequire}) and (\ref{eq:grequire}) are not as different as they might first appear, since $\bfy$ may be regarded as a parameter in the second equation; that is, we may write $\bfg(\bfy,\bfx)=\bfgy(\bfx)$.  So in either case, we seek functions -- either $\bff$ or $\bfgy$ -- mapping $\calC\rightarrow\calC$ whose self-composition is the identity.  If we find more than one such function for a given configuration space, we may let the first argument of $\bfg$, namely $\bfy$, determine which one we use for the second-order case.

\subsection{Permutation Representation}

Let $\Gamma$ be a second-order discrete dynamical system.  From the discussion above we have that $\Gamma$ is invertible if the associated function $\bfg(\bfy,\bfx)=\bfgy(\bfx)$ is a bijection $\bfgy:\calC\rightarrow\calC$; that is, $\Gamma$ is invertible if $\bfgy$ acts as a permutation on $\calC$ for all $\bfy\in\calC$.  We also have that $\Gamma$ is reversible if $\bfgy$ is a bijection and self-inverse; that is $\Gamma$ is reversible if $\bfgy^2$ is the identity, or $\bfgy$ acts as a permutation of order at most two on $\calC$ for all $\bfy\in\calC$ so that it is a product of disjoint transpositions.  The most obvious way to describe $\bfg$ is by explicitly listing the functions $\bfgy$ for all $\bfy\in\calC$.

\begin{example}
\label{ex:invertible}
A dynamical system of order $p=2$ with configuration space $\calC=\{0,1,2\}$ of cardinality $r=3$ may have dynamics described by the function
\begin{center}
\begin{tabular}{|cc|c|c|}
\hline  
$\bfx$ & $\bfy$ & $\bfgy(\bfx)$ & $\gy$ \\ 
\hline
$0$ & $0$ & $2$ &  \\
$1$ & $0$ & $0$ & $\sigma_1$ \\ 
$2$ & $0$ & $1$ &  \\ 
\hline
$0$ & $1$ & $0$ &  \\
$1$ & $1$ & $1$ & $e$ \\ 
$2$ & $1$ & $2$ &  \\ 
\hline
$0$ & $2$ & $0$ &  \\
$1$ & $2$ & $1$ & $e$ \\ 
$2$ & $2$ & $2$ &  \\
\hline
\end{tabular}
\end{center}
The last column of this table lists the elements of the symmetric group $S_3$ used for each value of $\bfy$; in this case, $\sigma_1=(021)$ is used for $\bfy=0$, and the identity permutation $e$ is used for $\bfy=1$ and $\bfy=2$.  Therefore the function $\bfg$ is uniquely specified by the sequence $\{\sigma_1,e,e\}$.  Because $\sigma_1$ has order three, this system is invertible but not reversible.
\end{example}

\begin{example}
\label{ex:reversible}
We consider the dynamical system of Example~\ref{ex:invertible}, but this time with dynamics described by the function
\begin{center}
\begin{tabular}{|cc|c|c|}
\hline  
$\bfx$ & $\bfy$ & $\bfgy(\bfx)$ & $\gy $ \\ 
\hline
$0$ & $0$ & $0$ &  \\
$1$ & $0$ & $2$ & $\tau_0$ \\ 
$2$ & $0$ & $1$ &  \\ 
\hline
$0$ & $1$ & $0$ &  \\
$1$ & $1$ & $1$ & $e$ \\ 
$2$ & $1$ & $2$ &  \\ 
\hline
$0$ & $2$ & $0$ &  \\
$1$ & $2$ & $1$ & $e$ \\ 
$2$ & $2$ & $2$ &  \\
\hline
\end{tabular}
\end{center}
This example differs from the previous one in that $\tau_0=(21)$ is used for $\bfy=0$.  Therefore this function $\bfg$ is specified by the sequence $\{\tau_0,e,e\}$.  Because $\tau_0$ has order two, this system is reversible.
\end{example}

From the above examples, we see that we can always represent $\bfg$ uniquely as a sequence of $r$ elements of the symmetric group $S_r$, each of them corresponding to a permutation $\bfgy$,
\[
\bfg=\left\{g_0,g_1,\dots,g_{r-1}\right\},
\]
where $g_j$ is now used to denote general elements of the permutation group $S_r$.  If all elements of this sequence are identical, so that $\bfgy$ does not depend on $\bfy$, we say that the associated dynamics $\Gamma$ are {\it trivial}; otherwise, they are {\it nontrivial}.

\section{Action Principle}

\begin{definition}
A path of length $T$ connecting states $\bfr_i$ and $\bfr_f$ is a sequence of states in configuration space, $\bfr=\{\bfr(t)\in\calC\; |\; 0\leq t\leq T\}$, beginning with state $\bfr(0)=\bfr_i$ and ending with state $\bfr(T)=\bfr_f$.
\end{definition}
The following definitions refer to second-order dynamics $\Gamma$ described by a function $\bfg$:
\begin{definition}
We say that a path connecting $\bfr(0)=\bfr_i$ and $\bfr(T)=\bfr_f$ is a physical path of length $T$ if either $T\in\{1,2\}$, or $T>2$ and $\bfr(t)=\bfg(\bfr(t-1),\bfr(t-2))$ for $2\leq t\leq T$.  Otherwise, $\bfr$ is said to be a non-physical path.
\end{definition}
That is, a path is physical if every sequence of three consecutive points on it, $\{\bfr(t-2), \bfr(t-1), \bfr(t)\}$ obeys the evolution equation, Eq.~(\ref{eq:so}).

Given any pair of states $\bfr_i,\bfr_f\in\calC$, let us further denote the set of all physical paths between $\bfr_i$ and $\bfr_f$ by $U(\bfr_i;\bfr_f)$, the set of all the non-physical paths between $\bfr_i$ and $\bfr_f$ by $V(\bfr_i;\bfr_f)$, and their union  by $W(\bfr_i;\bfr_f)=U(\bfr_i;\bfr_f)\cup V(\bfr_i;\bfr_f)$.
\begin{definition}
Given a Lagrangian function $\calL:\calC\times\calC\longrightarrow\mathbb{R}$, the action of a path $\bfr$ is defined by
\begin{equation} 
S(\bfr)=\sum_{t=0}^{T-1}\calL(\bfr(t),\bfr(t+1)).
\end{equation}
\end{definition}
\begin{definition}
We say that $\Gamma$ admits an action principle if, for any $\bfr_i,\bfr_f\in\calC$ such that $U(\bfr_i,\bfr_f)\neq\emptyset$, and for any $\bfr\in U(\bfr_i;\bfr_f)$ and $\bfr'\in V(\bfr_i;\bfr_f)$, it is true that
\begin{equation} \label{eq:act}
S(\bfr)<S(\bfr')
\end{equation}
for some Lagrangian function $\calL:\calC\times\calC\longrightarrow\mathbb{R}$.
\end{definition}
That is, the action on any physical path is required to be less than the action on any nonphysical path.  We do not require any relative ordering of the action on two physical paths.

The following Lemma will be central to our analysis:
\begin{lemma}
\label{lem:lem1}
Let $\bfr\in U(\bfr_i;\bfr_f)$ be a physical path of length $T$. If $\Gamma$ admits an action principle, then
\begin{equation}\label{eq:cond1}
\calL(\bfr(t),\bfr(t+1))+\calL(\bfr(t+1),\bfr(t+2))<\calL(\bfr(t),\bfr'(t+1))+\calL(\bfr'(t+1),\bfr(t+2))
\end{equation}
for all  $\bfr'(t+1)\in\calC$ such that 
\begin{equation}\label{eq:cond2}
\bfg(\bfr'(t+1),\bfr(t))\neq\bfg(\bfr(t+1),\bfr(t)).
\end{equation}
\end{lemma}
{\it Proof.} We first notice that 
\[
\bfr'=\{\bfr_i=\bfr(0),\bfr(1),\ldots,\bfr(t),\bfr'(t+1),\bfr(t+2)\ldots,\bfr(T)=\bfr_f\}
\]
is a non-physical path of length $T$. Since, by hypothesis, we know that $S(\bfr)-S(\bfr')<0$, we have that 
\begin{eqnarray*}
\lefteqn{S(\bfr)-S(\bfr')=}\\
& &\calL(\bfr(t),\bfr(t+1))+\calL(\bfr(t+1),\bfr(t+2))-\calL(\bfr(t),\bfr'(t+1))-\calL(\bfr'(t+1),\bfr(t+2))<0.
\end{eqnarray*}
which proves the claim.
\begin{flushright}
$\square$
\end{flushright}

\begin{theorem}
Invertible, second-order, nontrivial, discrete dynamical systems with a finite configuration space do not admit an action principle.
\end{theorem}
{\it Proof}.
Let $\Gamma$ be a second-order discrete dynamical system with associated function $\bfg$ represented by the sequence  of permutations $\{g_0,g_1,\dots,g_{r-1}\}$.  Because the dynamical system is nontrivial, we may suppose that there exist $i,j\in\calC$ such that $g_i\neq g_j$.  If we set $\rho=g_jg^{-1}_i\neq e$, we can write $g_j$ in terms of $g_i$ as 
\[
g_j=\rho g_i.
\]
Using the tabular representation developed in the previous section, we can describe $\bfgi$ and $\bfgj$ as follows:
\begin{center}
\begin{tabular}{|cc|c|c|}
\hline  
$\bfx$ & $\bfy$ & $\bfgi(\bfx)$ & $\gy$ \\ 
\hline
$0$ & $i$ & $\bfg_i(0)$ &  \\
$1$ & $i$ & $\bfg_i(1)$ &  \\ 
$2$ & $i$ & $\bfg_i(2)$ & $g_i$ \\ 
$\vdots$ & $\vdots$ & $\vdots$ &  \\
$r-1$ & $i$ & $\bfg_i(r-1)$ &  \\ 
\hline
\end{tabular}
\hspace{0.5in}
\begin{tabular}{|cc|c|c|}
\hline  
$\bfx$ & $\bfy$ & $\bfgj(\bfx)$ & $\gy$ \\ 
\hline
$0$ & $j$ & $\rho(\bfg_i(0))$ &  \\
$1$ & $j$ & $\rho(\bfg_i(1))$ &  \\ 
$2$ & $j$ & $\rho(\bfg_i(2))$ & $g_j$ \\ 
$\vdots$ & $\vdots$ & $\vdots$ &  \\
$r-1$ & $j$ & $\rho(\bfg_i(r-1))$ &  \\ 
\hline
\end{tabular}
\end{center}
In order to simplify the notation, we also define $\xi_k\equiv\bfg_i(k)$.

To illustrate the main idea behind the proof of the theorem, we first consider the case when $\rho$ is a transposition; that is, for some $h,k \in\calC$, we have that $\rho=(hk)$ (in cycle notation).  Then the tabular representation of $\bfgi$ and $\bfgj$ can be written as follows:
\begin{center}
\begin{tabular}{|cc|c|c|}
\hline  
$\bfx$ & $\bfy$ & $\bfgi(\bfx)$ & $\gy$ \\ 
\hline
$0$ & $i$ & $\xi_0$ &  \\
$1$ & $i$ & $\xi_1$ &  \\ 
$\vdots$ & $\vdots$ & $\vdots$ &  \\
$h$ & $i$ & $\xi_h$ & $g_i$  \\ 
$\vdots$ & $\vdots$ & $\vdots$ &  \\
$k$ & $i$ & $\xi_k$ &  \\ 
$\vdots$ & $\vdots$ & $\vdots$ &  \\
$r-1$ & $i$ & $\xi_{r-1}$ &  \\ 
\hline
\end{tabular}
\hspace{0.5in}
\begin{tabular}{|cc|c|c|}
\hline  
$\bfx$ & $\bfy$ & $\bfgj(\bfx)$ & $\gy$ \\ 
\hline
$0$ & $j$ & $\xi_0$ &  \\
$1$ & $j$ & $\xi_1$ &  \\ 
$\vdots$ & $\vdots$ & $\vdots$ &  \\
$h$ & $j$ & $\xi_k$ &  $g_j$\\ 
$\vdots$ & $\vdots$ & $\vdots$ &  \\
$k$ & $j$ & $\xi_h$ &  \\ 
$\vdots$ & $\vdots$ & $\vdots$ &  \\
$r-1$ & $j$ & $\xi_{r-1}$ &  \\ 
\hline
\end{tabular}
\end{center}
By analyzing the representation of $\bfg_i$, we notice that $\{h,i,\xi_h\}$ and $\{k,i,\xi_k\}$ are physical paths whereas $\{h,j,\xi_h\}$ and $\{k,j,\xi_k\}$ are not. Similarly, by looking at the representation of $\bfg_j$, we notice that $\{h,j,\xi_k\}$ and $\{k,j,\xi_h\}$ are physical paths whereas $\{h,i,\xi_k\}$ and $\{k,i,\xi_h\}$ are not. Therefore, if $\Gamma$ admitted an action principle, then by Lemma~(\ref{lem:lem1}), $\calL$ would satisfy the linear constraints
\begin{equation}
\begin{array}{c}\calL(h,i)+\calL(i,\xi_h)<\calL(h,j)+\calL(j,\xi_h) \\\calL(k,i)+\calL(i,\xi_k)<\calL(k,j)+\calL(j,\xi_k) \\\calL(h,j)+\calL(j,\xi_k)<\calL(h,i)+\calL(i,\xi_k) \\\calL(k,j)+\calL(j,\xi_h)<\calL(k,i)+\calL(i,\xi_h)\end{array}.
\end{equation}
Adding the above inequalities, we notice that all terms cancel. Since the inequalities  are strict, we arrive at a contradiction.

Now let us generalize to the case where $\rho\in S_n$.  The tabular representation of $\bfgi$ and $\bfgj$ may be written as follows:
\begin{center}
\begin{tabular}{|cc|c|c|}
\hline  
$\bfx$ & $\bfy$ & $\bfgi(\bfx)$ & $\gy$ \\ 
\hline
$0$ & $i$ & $\xi_0$ &  \\
$1$ & $i$ & $\xi_1$ &  \\ 
$2$ & $i$ & $\xi_2$ & $g_i$ \\ 
$\vdots$ & $\vdots$ & $\vdots$ &  \\
$r-1$ & $i$ & $\xi_{r-1}$ &  \\ 
\hline
\end{tabular}
\hspace{0.5in}
\begin{tabular}{|cc|c|c|}
\hline  
$\bfx$ & $\bfy$ & $\bfgj(\bfx)$ & $\gy$ \\ 
\hline
$0$ & $j$ & $\rho(\xi_0)$ &  \\
$1$ & $j$ & $\rho(\xi_1)$ &  \\ 
$2$ & $j$ & $\rho(\xi_2)$ & $g_j$ \\ 
$\vdots$ & $\vdots$ & $\vdots$ &  \\
$r-1$ & $j$ & $\rho(\xi_{r-1})$ &  \\ 
\hline
\end{tabular}
\end{center}

If we define $I_{\rho}=\{i\in\calC\ |\ \rho(\xi_i)\neq \xi_i\}$, then, for any $k\in I_{\rho}$, by Lemma~\ref{lem:lem1}, we obtain the following contraints for the Lagrangian function $\calL$:
\begin{equation}\label{eq:sys}
\begin{array}{c}\calL(k,i)+\calL(i,\xi_k)<\calL(k,j)+\calL(j,\xi_k)      
\\\calL(k,j)+\calL(j,\rho(\xi_k))<\calL(k,i)+\calL(i,\rho(\xi_k))\end{array}.
\end{equation}

If we add both sides of the (\ref{eq:sys}), after cancellations we get:
\begin{equation}\label{eq:ineq}
\calL(i,\xi_k)+\calL(j,\rho(\xi_k))<\calL(j,\xi_k)+\calL(i,\rho(\xi_k))
\end{equation}
where $k\in I_{\rho}$.

After noticing that 
\[\sum_{k\in I_{\rho}}\calL(\rho(i,\xi_k))=\sum_{k\in I_{\rho}}\calL(i,\xi_k)\]
for any $i\in\calC$, we can conclude that the sum of the left-hand sides of the inequalities (\ref{eq:ineq}) coincides with the sum of their right-hand sides.  Again, since the inequalities are strict, we arrive at a contradiction.
\begin{flushright}
$\square$
\end{flushright}

\begin{example}
Let us consider the discrete dynamical system $\Gamma$ described by the following associated function $\bfg:$
\begin{center}
\begin{tabular}{|cc|c|c|}
\hline  
$\bfx$ & $\bfy$ & $\bfgy(\bfx)$ & $\gy$ \\ 
\hline
$0$ & $0$ & $1$ &  \\
$1$ & $0$ & $0$ &  \\
$2$ & $0$ & $3$ & $(01)(23)$ \\ 
$3$ & $0$ & $2$ &  \\ 
\hline
$0$ & $1$ & $2$ &  \\
$1$ & $1$ & $3$ &  \\
$2$ & $1$ & $0$ & $(02)(13)$ \\ 
$3$ & $1$ & $1$ &  \\ 
\hline
$0$ & $2$ & $1$ &  \\
$1$ & $2$ & $0$ &  \\
$2$ & $2$ & $3$ & $(01)(23)$ \\ 
$3$ & $2$ & $2$ &  \\
\hline
$0$ & $3$ & $1$ &  \\
$1$ & $3$ & $0$ &  \\
$2$ & $3$ & $3$ & $(01)(23)$ \\ 
$3$ & $3$ & $2$ &  \\
\hline
\end{tabular}
\end{center}
If $\Gamma$ admitted a non-trivial action principle, then, by looking at $\bfg_0$ and $\bfg_1$, we would infer that there exists a Lagrangian $\calL$ verifying the following inequalities:
\begin{eqnarray}
\nonumber\calL(0,0)+\calL(0,1)<\calL(0,1)+\calL(1,1) \\ 
\nonumber\calL(1,0)+\calL(0,0)<\calL(1,1)+\calL(1,0) \\
\nonumber\calL(2,0)+\calL(0,3)<\calL(2,1)+\calL(1,3) \\
\nonumber\calL(3,0)+\calL(0,2)<\calL(3,1)+\calL(1,2) \\
\nonumber\calL(0,1)+\calL(1,2)<\calL(0,0)+\calL(0,2) \\
\nonumber\calL(1,1)+\calL(1,3)<\calL(1,0)+\calL(0,3) \\
\nonumber\calL(2,1)+\calL(1,0)<\calL(2,0)+\calL(0,0) \\
\nonumber\calL(3,1)+\calL(1,1)<\calL(3,0)+\calL(0,1) \\ \nonumber
\end{eqnarray}
If we add the inequalities, after cancellations we arrive at the contradictory  inequality $0<0$.
\end{example}

\section{Recovering an action}

We have proved that invertible second-order discrete dynamical systems with a finite configuration space $\mathcal{C}$ do not admit an action defined on the entire phase space $\mathcal{C}^2=\mathcal{P}$. If we either allow $\mathcal{C}$ to be infinite,  or if we restrict to a subset of $\mathcal{P}$, the existence of an action principle is not ruled out by the above argument. In this section we present two examples of such instances where an action principle can be recovered.

\subsection{Infinite alphabet}

Using the notation previously introduced, let us consider a configuration space $\mathcal{C}$ of infinite size. Let then $\Gamma$ be described by a function $\bfg$ that can be represented by a sequence of automorphisms $g_i:\mathbb{Z}\longrightarrow\mathbb{Z}$
\[\bfg=\{g_i\}_{i\in\mathbb{Z}}\]
\begin{definition}
 $\Gamma$ is said to be of finite type if there exists $i\in\mathbb{Z}^+$ such that $g_i$ contains some non trivial finite cycle.
 \end{definition} 
 The following lemma is direct consequence of Theorem (1):
 \begin{lemma}
If $\Gamma$ is of finite type, then it does not admit an action principle.
\end{lemma}

By the next example we would like to highlight the fact that, if $\Gamma$ is not of finite type, an action principle may exist.
\begin{example}
Consider the dynamics induced by $\bfg(y,x)=x-y$, and let 
\[\calL(x,y)=\frac{1}{3}[(x+y)-(x+y)^2]+(x-y)^2\]
with $x,y\in\mathbb{Z}$

For any $y'\neq y$, we have that 
\[\calL(x,y)+\calL(y,y-x)-[\calL(x,y')-\calL(y',y-x)]=-\frac{2}{3}[-1+2(y-y')](y-y').\]
We notice that the above expression is always negative since, if $y-y'>0$ then 
\[
-1+2(y-y')>0
\]
(since $y-y'\geq1$). By the same token, if $y-y<0'$, then
\[
-1+2(y-y')<0.
\]
Hence we can conclude that $\calL(x,y)$ is a Lagrangian for the dynamics defined by $\bfg(y,x)=x-y$.

The above action is, however, not unique. Indeed, let us assume a quadratic form for the Lagrangian $\calL(x,y)$, that is
\[
\calL(x,y)=\alpha x^2+\beta xy+\gamma y^2,
\]
with $\alpha,\beta,\gamma\in\mathbb{R}.$
For any $y'\neq y$, we require that 
\begin{equation}
\label{eq:inf}
\calL(x,y)+\calL(y,y-x)-[\calL(x,y')-\calL(y',y-x)]=(y-y')[y'(\alpha+\gamma)+y(\alpha+\beta+\gamma)]<0.
\end{equation}
It is easy to show that solutions to $(\ref{eq:inf})$ must obey following form:
\[\alpha+\beta=C\ \ \ \alpha+\beta+\gamma=-C,\]
where $C\in\mathbb{R}$ is a constant.
That is, $\calL(x,y)=\alpha x^2+\beta xy+\gamma y^2$ is an action for $\bfg$ for any $\alpha,\beta,\gamma\in\mathbb{R}$ such that 
\[2\alpha+\beta+2\gamma=0.\]
 
\end{example}
\subsection{Restricting the phase space}

Let $\Gamma$ be a second-order discrete dynamical system, $\bfg$ its associated function, and let 
$(x,y)\in\calC^2=\calP$  be a point in its phase space. Then, as we have discussed in an earlier section, the dynamics in phase space $\calP$ is represented by the map $\Gamma$:
\[\Gamma:\calP\longrightarrow \calP\]
\[(x,y)\mapsto(y,\bfg (y,x)).\]
We define the {\it orbit} originated at $s\in\calP$ as:
\[\mathcal{O}_{s}=\{\Gamma^t(s)\ |\ t\in\mathbb{Z}^+\},\]
and decompose the global dynamics in phase space into a {\it transient} part 
\[W=\{s\in\calP\ |\ \forall\ t>0 : \Gamma^t(s)\neq s\} \]
and a {\it cyclic} part 
\[Z=\calP\backslash W\]

Notice that, if $\Gamma$ is invertible, then the transient part of the orbit through any point $s\in\calP$ is empty, and therefore $W=\emptyset$. On the other hand, if $\Gamma$ is non-invertible, then there are points in $\calP$ whose orbits have a non-trivial transient part, and therefore $W\neq\emptyset$. In this latter case, since the No-Go Theorem does not apply to non-invertible systems, an action may exist and we have explicitly computed a Lagrangian $\calL$ in some of these cases (see Example \ref{ex:noninv} below).

If $\calL$ is a Lagrangian inducing an action for the non-invertible system $\Gamma$, then we can recover an `invertible' action by restricting $\calL$ to the truncated phase space $Z=\calP\backslash T$. It is important to notice that this fact does not contradict our main result. In order to clarify that, given a finite set $A$, let us first define
\[
A^{*}=\{\sigma:A^2\longrightarrow A^2 \ |\ \sigma(x,y)=(y,z),\ \sigma \in\mathcal{S}_{|A|^2}\}
\]
for any $x,y\in A$ and some $z\in A.$ Here $\mathcal{S}_N$ denotes the symmetric group on $N$ elements.
\begin{example}
Let $A=\{0,1\}$. Then $A^2=\{00,01,10,11\}$, and it is not difficult to show that, out of the $4!$ possible permutations of elements of $A^2$, only four of them obey the above conditions, namely 
$A^{*}=\{\sigma_i\}_{i=1,\dots,4}$, where $\sigma_{1}(x,y)=(y,x)$, $\sigma_2(x,y)=(y,1-x)$, $\sigma_3(x,y)=(y,x\oplus y)$, $\sigma_4(x,y)=(y,1-x\oplus y)$. Here $x\oplus y$ denotes addition modulo 2.
\end{example}

Now we can observe that, by the No-Go Theorem, $\Gamma$ cannot be recast as a second-order dynamical system on a finite configuration space. That is, there exists no alphabet $\calC$ such that $\Gamma\in \cal C^{*}$. An interesting consequence of this observation is that, every time an action for some non-invertible system $\Gamma$ is detected, we can recover some ``negative information'' on its limit cycle structure -- namely, the fact that $\Gamma$ cannot belong to $\calC^{*}$. 

The following is an example of a non-invertible system which admits an action principle. 
\begin{example}\label{ex:noninv}
Let $|\calC|=3$ and consider the second-order,  non-reversible dynamics associated with the following function \bfg:
\[\left.\begin{array}{|c|c|c|}\hline \bfx & \bfy & \bfgy(x) \\\hline 0 & 0 & 2 \\\hline 1 & 0 & 1 \\\hline 2 & 0 & 0 \\\hline 0 & 1 & 2 \\\hline 1 & 1 & 1 \\\hline 2 & 1 & 0 \\\hline 0 & 2 & 1 \\\hline 1 & 2 & 0 \\\hline 2 & 2 & 0 \\\hline \end{array}\right.\]

In phase space, the dynamics is given by a 7-cycle
\[00\to02\to21\to10\to01\to12\to20,\]
a 1-cycle
\[11\to11,\]
and a single transient state 
\[22\to 20.\]

We can compute the constraints for the Lagrangian with the help of the following diagrams, which represent all possible paths (\bfx(t),\bfx(t+1),\bfx(t+2)). More precisely, the left and the right corners of the diamonds represent states of the system at time $t$ and at time $t+2$., respectively. That is, they denote $\bfx(t)$ and $\bfx(t+2)$, respectively. Then, for each pair $(\bfx(t),\bfx(t+2))$ there are $|\calC|=3$ possible intermediate states $\bfx(t+1)$, which are represented by the column between $\bfx(t)$ and $\bfx(t+2)$. Finally, we adopt the convention that plain arrows denote physical paths, while wiggly arrows denote unphysical paths.

We begin with
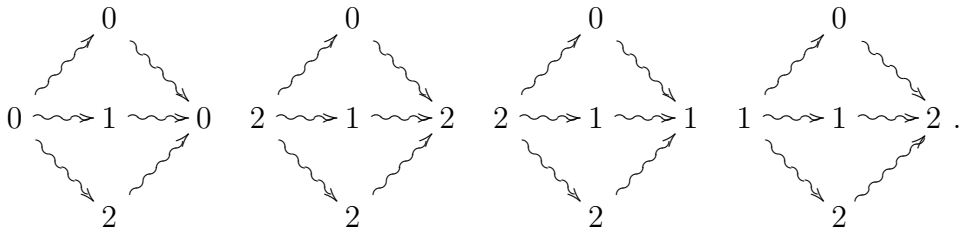
\begin{displaymath}
\xymatrix {
   & 0\ar@{~>}[dr]  &  \\
0   \ar@{~>}[ur]  \ar@{~>}[r] \ar@{~>}[dr]  & 1 \ar@{~>}[r]  & 0\\
 & 2 \ar@{~>}[ur]   & } \ \ 
\xymatrix {
   & 0\ar@{~>}[dr]  &  \\
2\ar@{~>}[dr] \ar@{~>}[r] \ar@{~>}[ur] & 1 \ar@{~>}[r]  & 2 \\
 & 2 \ar@{~>}[ur]  & }\ \ 
\xymatrix {
   & 0\ar@{~>}[dr]  &  \\
2\ar@{~>}[dr] \ar@{~>}[r] \ar@{~>}[ur] & 1 \ar@{~>}[r]  & 1 \\
 & 2 \ar@{~>}[ur]  & }\ \ 
\xymatrix {
   & 0  \ar@{~>}[dr]&  \\
1 \ar@{~>}[ur] \ar@{~>}[r] \ar@{~>}[dr] & 1 \ar@{~>}[r] & 2\ . \\
 & 2 \ar@{~>}[ur]  & }
\end{displaymath}
In the above diagrams, all paths are unphysical, hence these cases give rise to no constraints at all.

Next, we consider
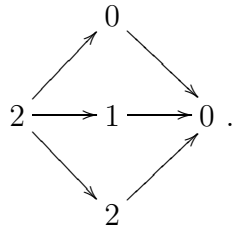
\begin{displaymath}
\xymatrix {
   & 0  \ar[dr]&  \\
2 \ar[ur] \ar[r] \ar[dr] & 1 \ar[r] & 0\ . \\
 & 2 \ar[ur]  & }
\end{displaymath}
In this case we notice that all paths are physical, therefore this case gives rise to no constraints as well.

Next, we consider consider the cases where there is only one physical path connecting $\bfx(t)$ with $\bfx(t+2)$:
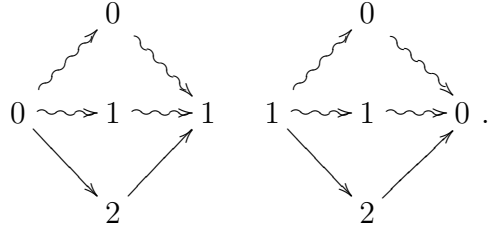
\begin{displaymath}
\xymatrix {
   & 0 \ar@{~>}[dr]   &  \\
0\ar[dr] \ar@{~>}[ur] \ar@{~>}[r]  & 1 \ar@{~>}[r] & 1 \\
 & 2 \ar[ur]  & }\ \ \ 
\xymatrix {
   & 0 \ar@{~>}[dr]  &  \\
1 \ar[dr] \ar@{~>}[r]  \ar@{~>}[ur]  & 1\ar@{~>}[r]  & 0\ . \\
 & 2 \ar[ur]  & }
\end{displaymath}
The inequality constraints arising from the above cases are
$$ \calL(0,2)+\calL(2,1)<\calL(0,1)+\calL(1,1) \ \ \ \ \ \ \ \ \ \ \ \ \ \calL(1,2)+\calL(2,0)<\calL(1,1)+\calL(1,0),$$
$$\calL(0,2)+\calL(2,1)<\calL(0,0)+\calL(0,1),  \ \ \ \ \ \ \ \ \ \ \ \ \calL(1,2)+\calL(2,0)<\calL(1,0)+\calL(0,0).$$

Finally, we have to consider the cases when there are two physical paths connecting $\bfx(t)$ with $\bfx(t+2)$:
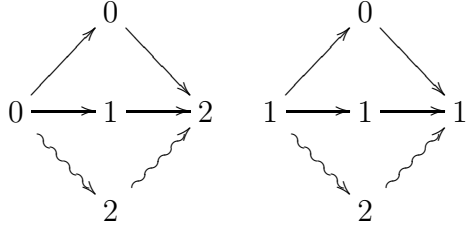
\begin{displaymath}
\xymatrix {
   & 0\ar[dr] &  \\
0\ar[r] \ar@{~>}[dr]  \ar[ur]& 1\ar[r] & 2 \\
 & 2 \ar@{~>}[ur]  & }\ \ \ 
 \xymatrix {
   & 0  \ar[dr]&  \\
1 \ar[ur] \ar[r] \ar@{~>}[dr]  & 1 \ar[r] & 1 \\
 & 2  \ar@{~>}[ur]  & }
\end{displaymath}
The inequality constraints arising from these cases are
$$\calL(0,0)+\calL(0,2)<\calL(0,2)+\calL(2,2),  \ \ \ \ \ \ \ \ \ \ \ \ \calL(1,0)+\calL(0,1)<\calL(1,2)+\calL(2,1), $$
$$\calL(0,1)+\calL(1,2)<\calL(0,2)+\calL(2,2),  \ \ \ \ \ \ \ \ \ \ \ \ \calL(1,1)+\calL(1,1)<\calL(1,2)+\calL(2,1).$$
\newtheorem{remark}{Remark}

More generally, note that if we denote by $N_p$ the number of physical paths, and by $N_n=|\calC|-N_p$ the number of unphysical paths, then the total number of inequalities is given by $N_pN_n=N_p(|\calC|-N_p)$.

As we have pointed out earlier, the No-Go Theorem does not preclude this non-reversible dynamics from having an action. In particular, if we remove the transient state 22 from the phase space, an action will be induced by a solution $\calL(x,y)$ to the system of inequalities given by
\[\left\{\begin{array}{c}\calL(0,2)+\calL(2,1)<\calL(0,1)+\calL(1,1) \\\calL(0,2)+\calL(2,1)<\calL(0,0)+\calL(0,1) \\\calL(1,2)+\calL(2,0)<\calL(1,1)+\calL(1,0) \\ \calL(1,2)+\calL(2,0)<\calL(1,0)+\calL(0,0) \\\calL(1,0)+\calL(0,1)<\calL(1,2)+\calL(2,1)  \\\calL(1,1)+\calL(1,1)<\calL(1,2)+\calL(2,1)\end{array}\right.\]

It can be seen that this system has an infinity of solutions (therefore the action is not unique). A particular solution is given, for instance, by
\begin{eqnarray*}
&\calL(0,0)=\calL(0,1)=\calL(0,2)=\calL(1,0)=\calL(1,1)=0&\\
&\calL(1,2)=2&\\
&\calL(2,0)=-3&\\
&\calL(2,1)=-1.&
\end{eqnarray*}
\end{example}

\section{Conclusions}

In this work we have discussed the question of the existence of action principles for invertible second-order discrete dynamical systems. We have proved that systems with a finite configuration space do not admit an action on the entire phase space. By weakening one of these two assumptions, that is assuming a configuration space of infinite size or working on a subset of the phase space, an action principle can be recovered and we have given examples of an action principle constructed in this way. 

Most of the motivations behind our work has come from the search for an action principle for finite-state cellular automata. Given their discrete nature, we decided to avoid any reference to differential structures, and have provided a genuinely discrete definition of variational principle by replacing the minimization of a continuous action with a system of linear inequalities obtained by minimizing a discrete action. In some sense, our result highlights both the similarities and the differences between continuous and discrete dynamical models. It seems to us that further investigations of these similarities and differences is warranted. After all, the elegance and power of continuous models in mechanics should not preclude the search for alternative, discrete models which could lead to useful applications, from both a theoretical and numerical-computational perspective. Future work might center on discrete analogs of other results in continuum mechanical systems, such as Noether's Theorem.

\section*{Acknowledgments}

This material is based on work supported in part by the US Army Research Laboratory and the US Army Research Office under Grant number W911NF-04-1-0334, in part by the US Air Force Office of Scientific Research under Grant number FA9550-04-1-0176, and in part by the DARPA QuIST Program under AFOSR Grant number F49620-01-1-0566.   The authors are grateful to Peter Love and Norman Margolus for their helpful comments and suggestions.

\end{document}